# 一種高效率金鑰擴展方法應用於安全憑證管理系統

Abel C. H. Chen*

中華電信研究院資通安全研究所

## 摘要

近年來，美國交通部也開始採用電機電子工程師學會(Institute of Electrical and Electronics Engineers, IEEE)的 1609 系列標準，建構安全憑證管理系統(Security Credential Management System, SCMS)作為美國車聯網的標準規範。其中，為提升車輛的隱私，在安全憑證管理系統中結合蝴蝶金鑰擴展方法(Butterfly Key Expansion, BKE)提供假名憑證(Pseudonym Certificate, PC)。然而，在 IEEE 1609.2.1 標準中，蝴蝶金鑰擴展方法主要是採用橢圓曲線密碼學(Elliptic Curve Cryptography, ECC)，擴展金鑰需要花費更多時間。有鑑於此，本研究提出一個原創的高效率金鑰擴展方法，並且通過數學原理論證此方法加解密的可行性、車輛的隱私、方法的效率。在實際環境以相同安全強度條件下比 IEEE 1609.2.1 2022 年標準的方法提升數千倍的效率。

# An Efficient Key Expansion Method Applied to Security Credential Management System

Abel C. H. Chen*

Information & Communications Security Laboratory, Chunghwa Telecom Laboratories

## Abstract

In recent years, U.S. Department of Transportation has adopts Institute of Electrical and Electronics Engineers (IEEE) 1609 series to build the security credential management system (SCMS) for being the standard of connected cars in U.S. Furthermore, a butterfly key expansion (BKE) method in SCMS has been designed to provide pseudonym certificates for improving the privacy of connected cars. However, the BKE method is designed based on elliptic curve cryptography (ECC) in the standard of IEEE 1609.2.1, but more execution time is required for key expansion. Therefore, this study proposes an original efficient key expansion method, and the mathematical principles have been proposed to prove the encryption/decryption feasibility, car privacy, and method efficiency. In a practical environment, the proposed method improves the efficiency of key expansion method in IEEE 1609.2.1-2022 with the same security strength thousands of times.

*Abel C. H. Chen E-mail: chchen.scholar@gmail.com
ORCID: 0000-0003-3628-3033

# 1. 前言

近年來，隨著智慧汽車、無人汽車的蓬勃發展，許多國際標準組織陸續對車聯網建立一系列的標準通訊標準、資訊安全標準、訊息格式標準等，包含電機電子工程師學會(Institute of Electrical and Electronics Engineers, IEEE)、歐洲電信標準協會(European Telecommunications Standards Institute, ETSI)、國際汽車工程學會(Society of Automotive Engineers International, SAE International)等標準組織(Ahmad et al., 2019; Molina-Masegosa et al., 2020)。然而，雖然已經許多標準制定好初稿和技術文件，但仍需各個國家政府介入規範要求車聯網相關終端設備和網路設備需符合哪些標準，才能讓車聯網相關企業依循和生產符合標準的終端設備和網路設備，最終才能打造出車聯網生態系統。

除此之外，車聯網要讓大眾廣泛接受和使用，仍需加強資訊安全和保護隱私。以車聯網資訊安全憑證管理系統為例，世界各國開始在公開金鑰基礎建設(Public Key Infrastructure, PKI)的基礎上結合非對稱式密碼學(如：RSA 演算法、橢圓曲線密碼學(Elliptic Curve Cryptography, ECC)演算法)和對稱式密碼學方法(如：高級加密標準(Advanced Encryption Standard, AES)演算法等)，建立符合車聯網環境的資訊安全憑證標準規範(Hammi, Monteuuis & Petit, 2022)。例如，近年來美國在 IEEE 標準為主體的基礎上提出安全憑證管理系統(Security Credential Management System, SCMS)(Walker, 2022; Barreto et al., 2021)，歐盟也在 ETSI 標準為主體的基礎提出協同智慧運輸系統憑證管理系統(Cooperative Intelligent Transportation System (C-ITS) Credential Management System, CCMS)(Directorate-General for Mobility and Transport, 2018; Berlato, Centenaro & Ranise, 2022)，制定符合北美環境和歐盟環境的車聯網資訊安全憑證標準規範。

其中，在 IEEE 1609.2.1 標準(Intelligent Transportation Systems Committee, 2022)中規範了安全憑證管理系統的系統架構，並且為了提升終端設備的隱私，提出了運用蝴蝶金鑰擴展方法(Butterfly Key Expansion, BKE)提供假名憑證(Pseudonym Certificate, PC)的作法(Ghosal & Conti, 2020)。並且，在蝴蝶金鑰擴展方法主要採用橢圓曲線密碼學的加解密和簽驗章方法(Simplicio et al, 2018)，以及在美國交通部在安全憑證管理系統概念性驗證(Proof of Concept, POC)主要採用國家標準暨技術研究院(National Institute of Standards and Technology)設計的橢圓曲線 NIST P-256 (Chen et al., 2019)來達到一定水平的安全強度。然而，雖然橢圓曲線密碼學可以提供較短的金鑰對，但在擴展金鑰所需的時間卻有可能花費比產製新的金鑰對所需的時間還多。

有鑑於此，本研究提出原創的高效率金鑰擴展方法，在符合蝴蝶金鑰擴展方法的精神上提出另一套密碼學方法來提供假名憑證。並且可以保證假名憑證使用上的隱私性，即使安全憑證管理系統中的伺服器間串供下也能保障終端設備的資訊不被破解。除此之外，本研究提出數學模型證明本研究的高效率金鑰擴展方法在加解密計算上的可行性，並且從原理上論證終端設備的隱私和方法的高效率。

本論文主要分為五個章節。第 2 節主要先從國際準介紹現有的車聯網資訊安全憑證標準規範，說明 IEEE 1609.2.1 安全憑證管理系統的系統架構及蝴蝶金鑰擴展方法。第 3 節將詳述本研究提出原創的高效率金鑰擴展方法，並且從原理上來論證本研究的方法。第 4 節將以實際環境來進行方法間的效率驗證，並且參考美國國家標準暨技術研究院所規範的安全強度標準(Barker, 2020)，在相同安全強度下進行方法效率比較。最後，第 5 節總結本研究的發現和討論未來方向。

# 2. IEEE 1609.2.1 安全憑證管理系統

本節將介紹 IEEE 1609.2.1 標準(Intelligent Transportation Systems Committee, 2022)的安全憑證管理系統，並且主要著重在與蝴蝶金鑰擴展方法有關的設備及流程進行說明。在 2.1 節中將介紹安全憑證管理系統架構，說明各個設備及其工作內容，在 2.2 節中將介紹蝴蝶金鑰擴展方法的具體流程，說明蝴蝶金鑰擴展方法的運作方式及設計精神。

## 2.1 安全憑證管理系統架構

在安全憑證管理系統中與蝴蝶金鑰擴展方法有關的設備主要有授權憑證中心(Authorization Certificate Authority, ACA)、註冊中心(Registration Authority)、以及終端設備(End Entity, EE)，如圖 1 所示。每個設備的工作說明如下：

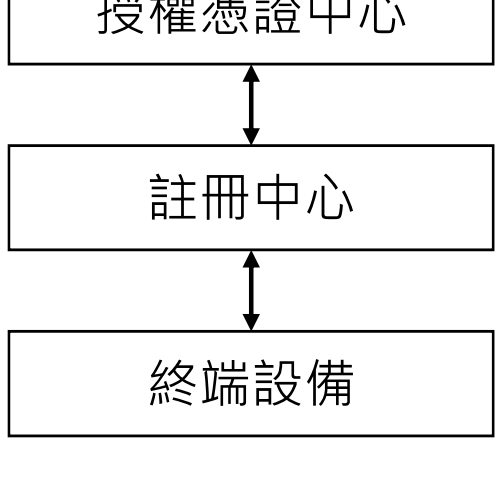

**圖 1　SCMS 架構(Intelligent Transportation Systems Committee, 2022)**

- 授權憑證中心可以簽發授權憑證(Authorization Certificate, AC)授予終端設備。並且為了保護終端設備的隱私性和安全性，授權憑證中心可以簽發假名憑證給終端設備，讓終端設備可以使用假名憑證進行通訊，避免終端設備經

常暴露其授權憑證。

- 註冊中心負責終端設備的註冊審核管理。
- 終端設備可以採用授權憑證或假名憑證在安全憑證管理系統中與其他設備進行安全通訊，通過憑證可以證實傳送內容是安全訊息。

## 2.2 蝴蝶金鑰擴展方法

蝴蝶金鑰擴展方法的主要流程包含終端設備產製毛蟲金鑰對，再由註冊中心基於毛蟲公鑰產製繭公鑰，並且由授權憑證中心基於繭公鑰產製蝴蝶公鑰，最後由終端設備產製繭私鑰和蝴蝶私鑰，如圖 2 所示。具體方法流程說明如下：

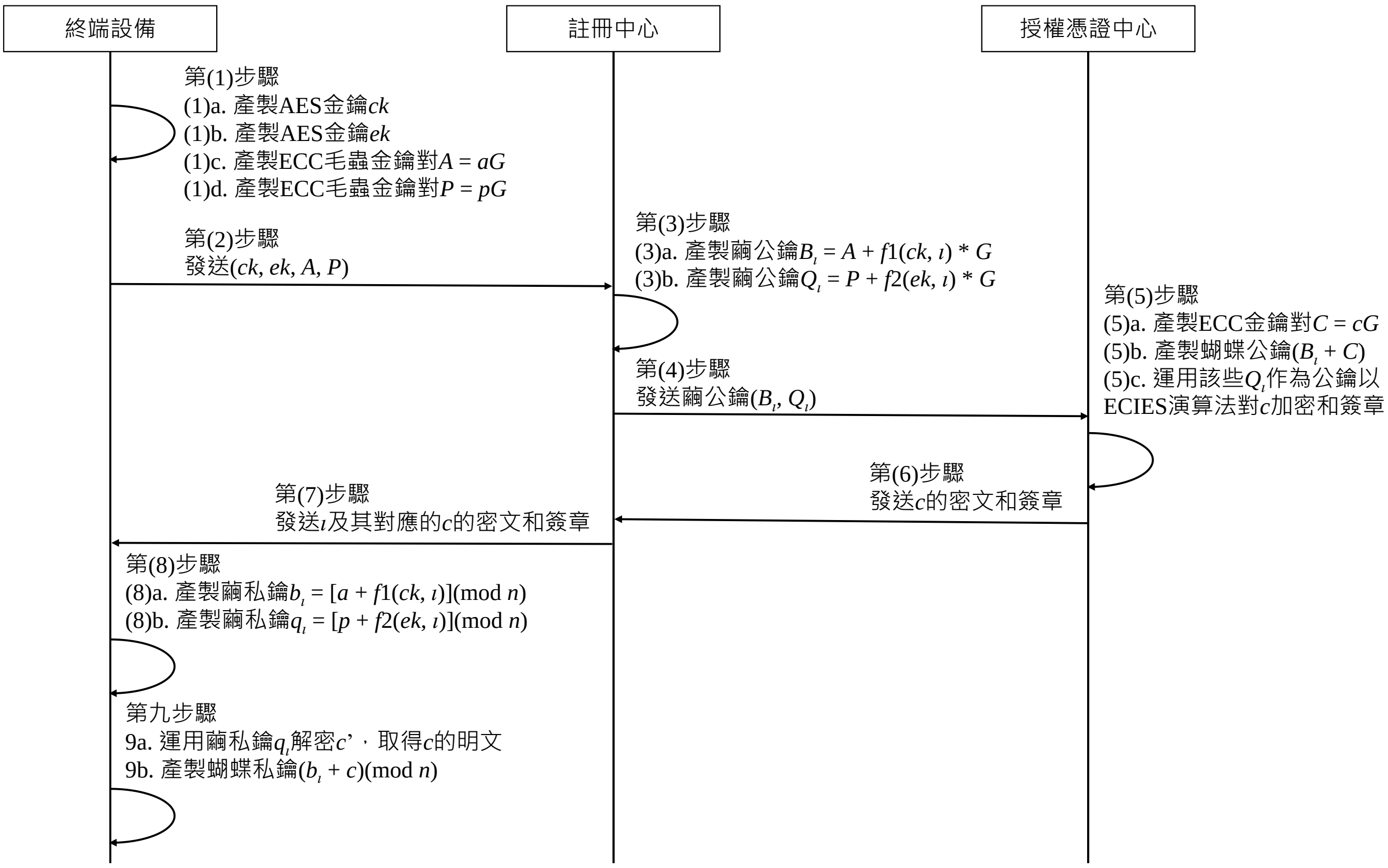

**圖 2　蝴蝶金鑰擴展方法(Intelligent Transportation Systems Committee, 2022)**

(1). 由終端設備產製 AES 金鑰及 ECC 金鑰對，具體方法流程包含有：

a. 產製 AES 金鑰 *ck*，作為簽章使用。其中，參數 *ck* 是對稱式金鑰。

b. 產製 AES 金鑰 *ek*，作為加密使用。其中，參數 *ek* 是對稱式金鑰。

c. 產製 ECC 金鑰對 $A = aG$，作為毛蟲金鑰對，簽章使用。其中，參數 $a$ 是私鑰、參數 $A$ 是公鑰、參數 $G$ 是橢圓曲線的基準點。

d. 產製 ECC 金鑰對 $P = pG$，作為毛蟲金鑰對，簽章使用。其中，參數 $p$ 是私鑰、參數 $P$ 是公鑰、參數 $G$ 是橢圓曲線的基準點。

(2). 由終端設備將產製的對稱式金鑰及毛蟲公鑰($ck, ek, A, P$)發送給註冊中心。

(3). 由註冊中心基於毛蟲公鑰產製繭公鑰，具體方法流程包含有：

a. 產製繭公鑰 $B_\iota = A + f1(ck, \iota) * G$。其中，參數 $\iota$ 是增量整數；函數 $f1$ 是基於 AES 加密演算法的擴展函數，可運用 AES 金鑰 $ck$ 加密參數 $\iota$ 值，得到整數密文，避免被攻擊者取得參數 ι 明文的情況下從繭公鑰推導出毛蟲公鑰。

b. 產製繭公鑰 $Q_\iota = P + f2(ek, \iota) * G$。其中，參數 $\iota$ 是增量整數；函數 $f2$ 是基於 AES 加密演算法的擴展函數，可運用 AES 金鑰 $ek$ 加密參數 $\iota$ 值，得到整數密文，避免被攻擊者取得參數 ι 明文的情況下從繭公鑰推導出毛蟲公鑰。

(4). 註冊中心將產製的繭公鑰($B_\iota, Q_\iota$)發送給授權憑證中心。

(5). 授權憑證中心基於繭公鑰產製蝴蝶公鑰，蝴蝶公鑰可以作為假名憑證的公鑰加密使用，具體方法流程包含有：

a. 產製 ECC 金鑰對 $C = cG$。其中，參數 $c$ 是私鑰、參數 $C$ 是公鑰、參數 $G$ 是橢圓曲線的基準點。

b. 產製蝴蝶公鑰($B_\iota + C$)。

c. 運用 $Q_\iota$ 作為公鑰以橢圓曲線整合加密機制(Elliptic Curve Integrated Encryption Scheme, ECIES)演算法對 $c$ 加密和簽章，其中 $c$ 的密文係 $c'$。

(6). 授權憑證中心發送密文 $c'$和簽章給註冊中心。

(7). 註冊中心發送 $\iota$ 值及其對應的密文 $c'$和簽章給終端設備。

(8). 終端設備根據 $\iota$ 值產製繭私鑰，具體方法流程包含有：

a. 產製繭私鑰 $b_\iota = [a + f1(ck, \iota)](\text{mod } n)$。其中，函數 $f1$ 是基於 AES 加密演算法的擴展函數，可運用 AES 金鑰 $ck$ 加密參數 $\iota$ 值，得到整數密文、參數 $n$ 是橢圓曲線的階。

b. 產製繭私鑰 $q_\iota = [p + f2(ek, \iota)](\text{mod } n)$。其中，函數 $f2$ 是基於 AES 加密演算法的擴展函數，可運用 AES 金鑰 $ek$ 加密參數 $\iota$ 值，得到整數密文、

參數 $n$ 是橢圓曲線的階。

(9). 終端設備使用繭私鑰 $q_ι$ 解密取得 $c$ 值，並且運用繭私鑰 $b_ι$ 和 $c$ 值產製蝴蝶私鑰，蝴蝶私鑰可作為假名憑證的私鑰簽章或解密使用，具體方法流程包含有：

a. 運用繭私鑰 $q_ι$ 解密密文 $c'$，取得明文 $c$。

b. 產製蝴蝶私鑰$(b_ι + c)(\mathrm{mod}\ n)$。其中，參數 $n$ 是橢圓曲線的階。

蝴蝶金鑰擴展方法可以保障終端設備的隱私，其理論基礎說明如下：

- 對註冊中心而言，因為註冊中心不知道明文 $c$ 值和 $C$ 值，所以無法從蝴蝶公鑰推導對應的繭公鑰。此外，註冊中心亦不知毛蟲私鑰，所以也無法推導繭私鑰和蝴蝶私鑰。

- 對授權憑證中心而言，因為授權憑證中心不知道擴展函數加密後的 $ι$ 值，所以無法從繭公鑰推導對應的毛蟲公鑰。此外，授權憑證中心亦不知毛蟲私鑰，所以也無法推導繭私鑰和蝴蝶私鑰。

## 2.3 蝴蝶金鑰擴展方法的效率限制

由於 IEEE 1609.2.1 標準的蝴蝶金鑰擴展方法主要採用橢圓曲線密碼學產製金鑰對，所以在擴展金鑰花費的時間將會比產製新的金鑰對的時間還長。詳細理論說明如下：

- 產製新的金鑰對：以 $A = aG$ 為例。其中，產生隨機整數 $a$ 的時間很短，所以時間主要花費在計算 $aG$。因此，雖然在數學表示上是乘法 $aG$，但橢圓曲線密碼學的主要限制在於計算累加了 $a$ 次的橢圓曲線的基準點 $G$，每次累加都會映射到橢圓曲線上的另一個點，與一般的整數相乘計算並不相同。

- 從毛蟲公鑰擴展成繭公鑰：以 $B_ι = A + f1(ck, ι) * G$ 為例。其中，擴展函數 $f1(ck, ι)$可以一個整數，所以計算 $f1(ck, ι) * G$ 的時間接近於產製新的金鑰對的時間，並且計算 $f1(ck, ι)$的時間高於產生隨機整數 $a$ 的時間，以及還需要與 $A$ 做計算。因此，從毛蟲公鑰擴展成繭公鑰的時間高於產製新的金鑰對的時間。

- 從繭公鑰擴展成蝴蝶公鑰：以蝴蝶公鑰$(B_ι + C)$為例。其中，需先產製 ECC 金鑰對 $C = cG$，所以計算蝴蝶公鑰$(B_ι + C)$的時間高於產製新的金鑰對的時間。因此，從繭公鑰擴展成蝴蝶公鑰的時間高於產製新的金鑰對的時間。

有鑑於橢圓曲線密碼學在金鑰擴展上的限制，本研究提出一套高效率金鑰擴展方法，在相同安全強度條件下提供更快速的金鑰擴展方法。

# 3. 原創的高效率金鑰擴展方法

本節將介紹本研究原創的高效率金鑰擴展方法，在 3.1 節介紹高效率金鑰擴展方法的具體流程和作法，在 3.2 節用數學模型證明本研究提出的高效率金鑰擴展方法的可行性，在 3.3 節以一個簡單實例說明本研究提出的高效率金鑰擴展方法。

## 3.1 方法流程

本研究提出原創的高效率金鑰擴展方法，主要包含終端設備產製 RSA 毛蟲金鑰對(即毛蟲私鑰 $s$ 與毛蟲公鑰 $S$)和擴展值 $\alpha$ 與 $\beta$，再由註冊中心基於 RSA 毛蟲公鑰 $S$ 與擴展值 $\alpha$ 產製 RSA 繭公鑰 $J_i$，並且由授權憑證中心基於 RSA 繭公鑰 $J_i$ 與擴展值 $\beta$ 產製 RSA 蝴蝶公鑰 $H_i$，最後由終端設備以蟲私鑰 $s$ 作為繭私鑰和蝴蝶私鑰，如圖 3 所示。具體方法流程說明如下：

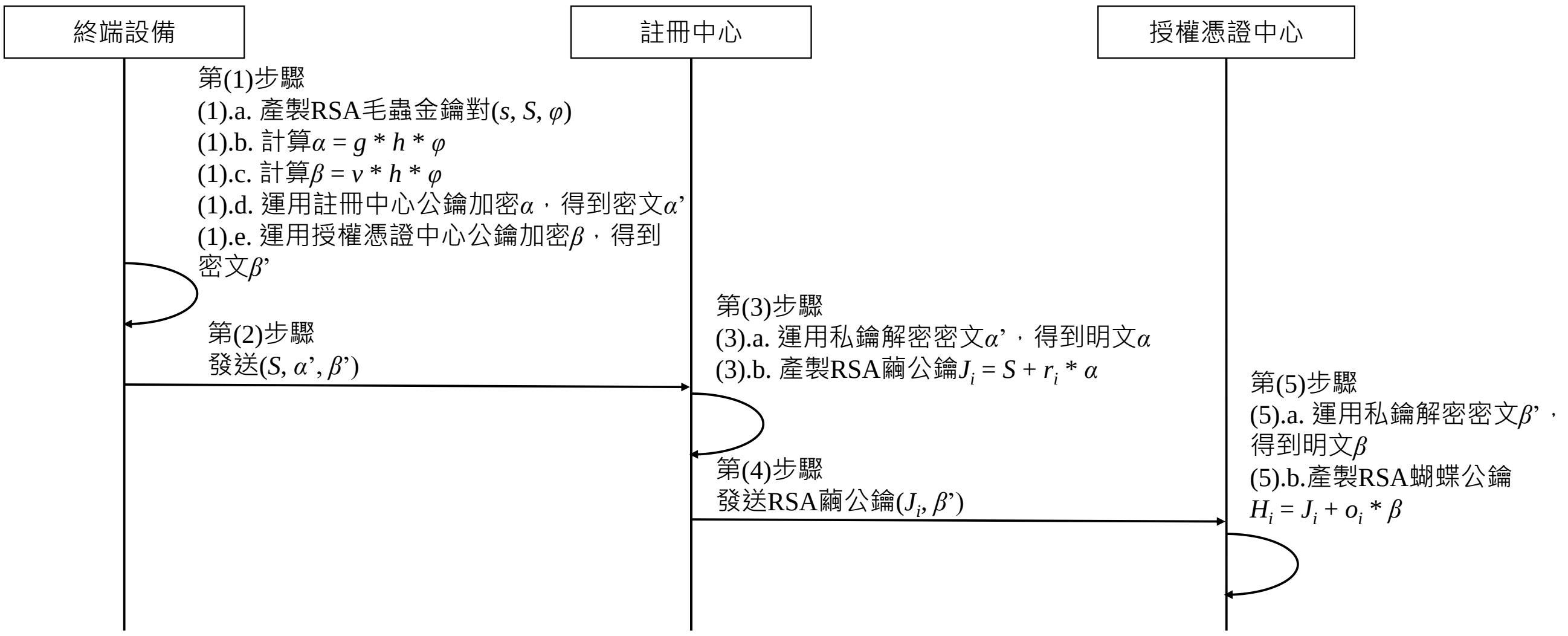

**圖 3　本研究原創的高效率金鑰擴展方法**

(1). 由終端設備產製 RSA 毛蟲金鑰對(即毛蟲私鑰 $s$ 與毛蟲公鑰 $S$)和擴展值 $\alpha$ 與 $\beta$，具體方法流程包含有：

a. 產製 RSA 毛蟲金鑰對：運用 RSA 演算法<u>隨機挑選兩個足夠大的質數分別為 $\rho$ 和 $\zeta$，兩質數相乘後可得 $N = \rho * \zeta$，並且從 $N$ 推導 $\rho$ 或 $\zeta$ 是離散對數問題(discrete logarithm problem)，將此數學原理作為**引理 1**</u>。計算 RSA 金鑰對的階 $\varphi = (\rho - 1)(\zeta - 1)$，再取得隨機挑選整數毛蟲私鑰 $s$ 和毛蟲公鑰 $S$，並且可以符合條件 $sS \pmod{\varphi} = 1$。

b. 計算擴展值 $\alpha$：產生隨機挑選兩個足夠大的質數分別為 $g$ 和 $h$，擴展值

$\alpha = g * h * \varphi$。

c. 計算擴展值 $\beta$：產生隨機挑選足夠大的質數 $v$，擴展值 $\beta = v * h * \varphi$。

d. 運用註冊中心的長期公鑰加密 $\alpha$，得到密文 $\alpha'$。

e. 運用授權憑證中心的長期公鑰加密 $\beta$，得到密文 $\beta'$。

(2). 由終端設備將產製的毛蟲公鑰及擴展值密文($S, \alpha', \beta'$)發送給註冊中心。

(3). 由註冊中心基於 RSA 毛蟲公鑰 $S$ 與擴展值 $\alpha$ 產製 RSA 繭公鑰 $J_i$，具體方法流程包含有：

a. 運用註冊中心的長期私鑰解密密文 $\alpha'$，得到明文 $\alpha$。

b. 產製 RSA 繭公鑰 $J_i = S + r_i * \alpha$。其中，參數 $r_i$ 是隨機整數，可以避免被攻擊者從繭公鑰推導出毛蟲公鑰。

(4). 註冊中心將產製的繭公鑰及擴展值密文($J_i, \beta'$)發送給授權憑證中心。

(5). 授權憑證中心基於 RSA 繭公鑰 $J_i$ 與擴展值 $\beta$ 產製 RSA 蝴蝶公鑰 $H_i$，蝴蝶公鑰可以作為假名憑證的公鑰加密使用，具體方法流程包含有：

a. 運用授權憑證中心的長期私鑰解密密文 $\beta'$，得到明文 $\beta$。

b. 產製 RSA 蝴蝶公鑰 $H_i = J_i + o_i * \beta$。其中，參數 $o_i$ 是隨機整數，可以避免被攻擊者從蝴蝶公鑰推導出繭公鑰。

## 3.2 原理證明

本節將先從加解密的數學原理證明本研究提出的金鑰擴展方法在加解密計算的可行性，再從保障終端設備的隱私提出證明，最後再論證本研究提出的金鑰擴展方法的效率。

### 3.2.1 加解密的數學原理證明

本節對 RSA 加解密數學原理和擴展後金鑰加解密進行數學原理證明。

- RSA 金鑰對加解密數學原理：符合 $sS \pmod{\varphi} = 1$ 條件時，明文 $x$ 可用公鑰 $S$ 加密得到密文 $x^S$，再用私鑰 $s$ 解密得到明文$(x^S)^s \pmod{N} = x^{sS} \pmod{N} = x$，將此數學原理作為**引理 2**。其中，參數 $s$ 是私鑰、參數 $S$ 是公鑰、參數 $\varphi$ 是 RSA 金鑰對的階。

- RSA 繭公鑰加解密數學原理：本研究的 RSA 繭公鑰是 $J_i = S + r_i * \alpha$，RSA 繭私鑰與毛蟲私鑰相同是 $s$。因此，明文 $x$ 可用 RSA 繭公鑰 $J_i$ 加密得到密文 $x^{J_i}$，再用私鑰 $s$ 解密得到明文$(x^{J_i})^s$ (mod $N$)。其中，通過公式(1)可以證明 $sJ_i$ (mod $\varphi$) = $sS$ (mod $\varphi$) = 1，並且根據**引理 2** 可證明公式(2)，成功解密密文 $x^{J_i}$，得到明文 $x^{sJ_i}$ (mod $N$) = $x^{sS}$ (mod $N$) = $x$。

$$\begin{aligned} sJ_i \pmod{\varphi} &= [s * (S + r_i * \alpha)] \pmod{\varphi} \\ &= [s * (S + r_i * (g * h * \varphi))] \pmod{\varphi} \\ &= [sS + s * r_i * (g * h * \varphi)] \pmod{\varphi} \\ &= sS \pmod{\varphi} = 1 \end{aligned} \tag{1}$$

$$x^{sJ_i} \pmod{N} = x^{sS} \pmod{N} = x \tag{2}$$

- RSA 蝴蝶公鑰加解密數學原理：本研究的 RSA 蝴蝶公鑰是 $H_i = J_i + o_i * \beta$，RSA 蝴蝶私鑰與毛蟲私鑰相同是 $s$。因此，明文 $x$ 可用 RSA 蝴蝶公鑰 $H_i$ 加密得到密文 $x^{H_i}$，再用私鑰 $s$ 解密得到明文$(x^{H_i})^s$ (mod $N$)。其中，通過公式(3)可以證明 $sH_i$ (mod $\varphi$) = $sS$ (mod $\varphi$) = 1，並且根據**引理 2** 可證明公式(4)，成功解密密文 $x^{H_i}$，得到明文 $x^{sH_i}$ (mod $N$) = $x^{sS}$ (mod $N$) = $x$。

$$\begin{aligned} sH_i \pmod{\varphi} &= [s * (J_i + o_i * \beta)] \pmod{\varphi} \\ &= [s * (S + r_i * (g * h * \varphi) + o_i * (v * h * \varphi))] \pmod{\varphi} \\ &= [sS + s * h * \varphi * (r_i * g + o_i * v)] \pmod{\varphi} \\ &= sS \pmod{\varphi} = 1 \end{aligned} \tag{3}$$

$$x^{sH_i} \pmod{N} = x^{sS} \pmod{N} = x \tag{4}$$

### 3.2.2 隱私證明

本研究提出的高效率金鑰擴展方法可以保障終端設備的隱私，其理論基礎說明如下：

- 對註冊中心而言，因為註冊中心不知道明文 $\beta$ 值，所以無法從蝴蝶公鑰推導對應的繭公鑰。此外，註冊中心亦不知毛蟲私鑰 $s$ 和 $\varphi$ 值，而且從**引理 1** 可

以證明無法從擴展值 $\alpha$ 推導出 $\varphi$ 值，所以也無法推導繭私鑰和蝴蝶私鑰。

- 對授權憑證中心而言，因為授權憑證中心不知道明文 $\alpha$ 值，所以無法從繭公鑰推導對應的毛蟲公鑰。此外，授權憑證中心亦不知毛蟲私鑰 $s$ 和 $\varphi$ 值，而且從**引理 1** 可以證明無法從擴展值 $\beta$ 推導出 $\varphi$ 值，所以也無法推導繭私鑰和蝴蝶私鑰。

- 除此之外，本研究為避免在註冊中心和授權憑證中心串供的情況下曝露 $\varphi$ 值，本研究的擴展值分別是 $\alpha = g * h * \varphi$ 和 $\beta = v * h * \varphi$。即使註冊中心和授權憑證中心串供同時獲得 $\alpha$ 和 $\beta$，可以求取最大公因數是 $h * \varphi$；然而，從**引理 1** 可以證明 $h * \varphi$ 求取 $\varphi$ 值，仍是一個離散對數問題。

### 3.2.3 效率證明

本研究提出的高效率金鑰擴展方法可以具有較高的效率，其理論基礎說明如下：

- 從毛蟲公鑰擴展成繭公鑰：以 $J_i = S + r_i * \alpha$ 為例。因為，毛蟲公鑰 $S$、隨機整數 $r_i$、以及擴展值 $\alpha$ 都是整數，所以在金鑰擴展時只是做一般的整數乘法計算，可以具有較高的效率。

- 從繭公鑰擴展成蝴蝶公鑰：以 $H_i = J_i + o_i * \beta$ 為例。因為，繭公鑰 $J_i$、隨機整數 $o_i$、以及擴展值 $\beta$ 都是整數，所以在金鑰擴展時只是做一般的整數乘法計算，可以具有較高的效率。

- 除此之外，本研究提出的高效率金鑰擴展方法僅需一組 RSA 毛蟲金鑰對(即毛蟲私鑰 $s$ 與毛蟲公鑰 $S$)和擴展值 $\alpha$ 與 $\beta$ 即可擴展成 RSA 繭公鑰和 RSA 蝴蝶公鑰。並且，計算擴展值時只是做一般的整數乘法計算，可以具有較高的效率。

- 本研究提出的金鑰擴展方法在效率的優勢主要包含：

  (1). 不需要產製多組的 AES 對稱式金鑰。

  (2). 不需要產製多組的非對稱式金鑰對，僅需一組 RSA 金鑰對。

  (3). 不需要執行擴展函數 $f1$ 和 $f2$ 進行 AES 演算法加密。

  (4). 從毛蟲公鑰擴展成繭公鑰和從繭公鑰擴展成蝴蝶公鑰都只是做一般的整數乘法計算，可以具有較高的效率。

(5). 不需要從註冊中心和授權憑證中心下載資料回終端設備。

(6). 不需要擴展私鑰。

## 3.3 案例說明

本節提供以 10 比特(bits)長度之整數的計算例進行說明，但實際上線系統建議應設定為 3072 比特長度以上之整數，以保障安全強度。具體案例說明如下：

(1). 由終端設備產製 RSA 毛蟲金鑰對和擴展值 $\alpha$ 與 $\beta$，具體方法流程包含有：

a. 產製 RSA 毛蟲金鑰對：運用 RSA 演算法隨機挑選兩個足夠大的質數分別為 $\rho$ = 991 和 $\zeta$ = 827，兩質數相乘後可得 $N = \rho * \zeta = 819557$。計算 RSA 金鑰對的階 $\varphi = (\rho - 1)(\zeta - 1) = 990 * 826 = 817740$，再取得隨機挑選整數毛蟲私鑰 $s$ = 84983 和毛蟲公鑰 $S$ = 65567，並且可以符合條件 $sS \pmod{\varphi} = 5572080361 \pmod{817740} = 1$。

b. 計算擴展值 $\alpha$：產生隨機挑選兩個足夠大的質數分別為 $g$ = 937 和 $h$ = 599，擴展值 $\alpha = g * h * \varphi = 937 * 599 * 817740 = 458967205620$。

c. 計算擴展值 $\beta$：產生隨機挑選足夠大的質數 $v$ = 983，擴展值 $\beta = v * h * \varphi = 983 * 599 * 817740 = 481499213580$。

d. 運用註冊中心的長期公鑰加密 $\alpha$，得到密文 $\alpha'$。在此計算例中，長期公鑰加密的密文以代數 $\alpha'$表示。

e. 運用授權憑證中心的長期公鑰加密 $\beta$，得到密文 $\beta'$。在此計算例中，長期公鑰加密的密文以代數 $\beta'$表示。

(2). 由終端設備將產製的毛蟲公鑰及擴展值密文$(S, \alpha', \beta') = (65567, \alpha', \beta')$發送給註冊中心。

(3). 由註冊中心基於 RSA 毛蟲公鑰與擴展值 $\alpha$ 產製 RSA 繭公鑰 $J_i$，並且假設產生兩個繭公鑰 $J_i$ 分別是 $J_1$ 和 $J_2$，具體方法流程包含有：

a. 運用註冊中心的長期私鑰解密密文 $\alpha'$，得到明文 $\alpha$ = 458967205620。

b. 產製 RSA 繭公鑰 $J_1 = S + r_1 * \alpha = 65567 + 2 * 458967205620 = 917934476807$。其中，參數 $r_1$ 是隨機整數，在此例中是 2。

產製 RSA 繭公鑰 $J_2 = S + r_2 * \alpha = 65567 + 3 * 458967205620 = 1376901682427$。其中，參數 $r_2$ 是隨機整數，在此例中是 3。

(4). 註冊中心將產製的繭公鑰及擴展值密文($J_i$, $\beta$') = ({917934476807, 1376901682427}, $\beta$')發送給授權憑證中心。

(5). 授權憑證中心基於 RSA 繭公鑰 $J_i$ 與擴展值 $\beta$ 產製 RSA 蝴蝶公鑰 $H_i$，並且假設產生兩個蝴蝶公鑰 $H_i$ 分別是 $H_1$ 和 $H_2$，具體方法流程包含有：

a. 運用授權憑證中心的長期私鑰解密密文 $\beta$'，得到明文 $\beta$ = 481499213580。

b. 產製 RSA 蝴蝶公鑰 $H_1 = J_1 + o_1 * \beta$= 917934476807 + 5 * 481499213580 = 3325430544707。其中，參數 $o_1$ 是隨機整數，在此例中是 5。
產製 RSA 蝴蝶公鑰 $H_2 = J_2 + o_2 * \beta$ = 1376901682427 + 7 * 4747396177487 = 917934476807。其中，參數 $o_2$ 是隨機整數，在此例中是 7。

產生 RSA 蝴蝶公鑰後，可將 RSA 蝴蝶公鑰作為假名憑證的公鑰進行資料加密。例如，明文 x = 101，採用 RSA 蝴蝶公鑰 $H_1$ = 3325430544707 進行加密得到密文 $101^{3325430544707}$。終端設備收到密文後，可以用蝴蝶私鑰(也就是毛蟲私鑰 $s$ = 84983)進行解密得到明文$[(101^{3325430544707})^{84983}]$(mod 819557) = 101。

# 4. 實驗分析與討論

在本節中將採用美國國家標準暨技術研究院所規範的安全強度標準(Barker, 2020)在實際環境來驗證不同方法做金鑰擴展的效率。在 4.1 節中描述實驗環境和安全強度標準，在 4.2 中描述和探討實驗結果。

## 4.1 實驗環境

為實際驗證本研究提出的高效率金鑰擴展方法之效率，本研究採用一台 Windows 10 企業版的電腦模擬註冊中心和授權憑證中心，驗證從毛蟲公鑰擴展成繭公鑰和從繭公鑰擴展成蝴蝶公鑰的時間。其中，實驗使用的軟硬體詳細規格是 CPU Intel(R) Core(TM) i7-10510U、記憶體 8 GB、OpenJDK 18.0.2.1、以及函式庫 Bouncy Castle Release 1.72。

為了比較在不同安全強度下的方法效率，本研究採用美國國家標準暨技術研究院所規範的安全強度標準(Barker, 2020)，如表 1 所示。可將安全強度分為五個等級，並每個等級對應密碼學的要求來實作。除此之外，本研究總共設計 4 個實驗情境，將本研究提出的高效率金鑰擴展方法(採用 RSA 密碼學)與 IEEE 1609.2.1

的 2022 年版標準方法(採用 ECC 密碼學)(Intelligent Transportation Systems Committee, 2022)各執行 1000 次進行比較，詳細說明如下。

- 實驗 1：從毛蟲公鑰擴展成繭公鑰，並只擴展成 1 個繭公鑰。
- 實驗 2：從繭公鑰擴展成蝴蝶公鑰，並只擴展成 1 個蝴蝶公鑰。
- 實驗 3：從毛蟲公鑰擴展成繭公鑰，並擴展成 20 個繭公鑰。
- 實驗 4：從繭公鑰擴展成蝴蝶公鑰，並擴展成 20 個蝴蝶公鑰。

**表 1　安全強度標準(Barker, 2020; Chen et al., 2019)**

| 安全強度(Security Strength) | RSA | ECC |
| --- | --- | --- |
| 80 | 1024 | 160<br>採用 NIST P-192 實作 |
| 112 | 2048 | 224<br>採用 NIST P-224 實作 |
| 128 | 3072 | 256<br>採用 NIST P-256 實作 |
| 192 | 7680 | 384<br>採用 NIST P-384 實作 |
| 256 | 15360 | 512<br>採用 NIST P-521 實作 |

其中，美國交通部在安全憑證管理系統概念性驗證(POC)時，設計一週發放 20 張假名憑證，所以本研究中考慮擴展成 20 個繭公鑰和 20 個蝴蝶公鑰的實驗來進行比較。

## 4.2 實驗結果與討論

本節共分為 3 個小節，前 2 個小節各別討論每個實驗的實驗結果，4.2.3 節總結 4 個實驗的發現和深入探討實驗結果。

### 4.2.1 實驗 1 和實驗 2：只擴展成 1 個繭公鑰、1 個蝴蝶公鑰

本節將整理實驗 1 和實驗 2 的實驗結果，觀察從毛蟲公鑰擴展成繭公鑰和從繭公鑰擴展成蝴蝶公鑰各擴展 1 個公鑰的結果，並且統計每個實驗中的 1,000 資料用“平均值(標準差)”的方式呈現每個方法的金鑰擴展時間，單位採用微秒 (microsecond)。

實驗 1 的金鑰擴展時間整理如表 2 所示。其中，安全強度 80 時，本研究方法花費的金鑰擴展平均時間為 11.861 微秒，而 IEEE 1609.2.1-2022 方法花費的金鑰擴展平均時間為 55589.152 微秒。本研究方法較 IEEE 1609.2.1-2022 方法在效率上有顯著提升的原因主要如 3.2.3 節所述：從毛蟲公鑰擴展成繭公鑰時，IEEE 1609.2.1-2022 方法需計算兩個繭公鑰，並且這兩個繭公鑰都需要計算 AES 演算法基礎的擴展函數對擴展值加密，所以需要花費大量的運算時間在擴展值加密及其累加擴展值密文倍數的橢圓曲線基準點 $G$。

除此之外，在不同安全強度的等級下，本研究方法在金鑰擴展時主要只是做一般整數計算，所以金鑰擴展時間沒有顯著差異。而 IEEE 1609.2.1-2022 方法在金鑰擴展時主要時間花費在累加擴展值密文倍數的橢圓曲線基準點 $G$，所以取決於擴展值密文值的大小；在擴展值密文值的大小無顯著差異的情況下，IEEE 1609.2.1-2022 方法在不同安全強度的等級下，金鑰擴展時間沒有顯著差異。

**表 2　實驗 1 金鑰擴展時間(單位：微秒)**

| 安全強度 | 本研究方法 | IEEE 1609.2.1-2022 |
|---|---|---|
| 80 | 11.861 (14.086) | 55589.152 (24823.229) |
| 112 | 11.613 (23.569) | 54905.265 (9530.397) |
| 128 | 13.008 (16.743) | 54347.954 (9686.334) |
| 192 | 10.163 (28.162) | 59667.727 (10590.645) |
| 256 | 10.565 (5.117) | 57025.756 (11479.287) |

實驗 2 的金鑰擴展時間整理如表 3 所示。其中，安全強度 80 時，本研究方法花費的金鑰擴展平均時間為 12.337 微秒，而 IEEE 1609.2.1-2022 方法花費的金鑰擴展平均時間為 18172.760 微秒。本研究方法較 IEEE 1609.2.1-2022 方法在效率上有顯著提升的原因主要如 3.2.3 節所述：從繭公鑰擴展成蝴蝶公鑰時，IEEE 1609.2.1-2022 方法需產製一組 ECC 金鑰對 $C = cG$，所以需要花費大量的運算時間在計算 $c$ 倍數的橢圓曲線基準點 $G$。不過由於在計算蝴蝶公鑰時，IEEE

1609.2.1-2022 方法需產製一組 ECC 金鑰對，所以金鑰擴展時間較實驗 1 時短。

除此之外，在不同安全強度的等級下，本研究方法在金鑰擴展時主要只是做一般整數計算，所以金鑰擴展時間沒有顯著差異。而 IEEE 1609.2.1-2022 方法在金鑰擴展時主要時間花費在計算 *c* 倍數的橢圓曲線基準點 *G*，所以取決於 *c* 值的大小；在 *c* 值的大小無顯著差異的情況下，IEEE 1609.2.1-2022 方法在不同安全強度的等級下，金鑰擴展時間沒有顯著差異。

**表 3　實驗 2 金鑰擴展時間(單位：微秒)**

| 安全強度 | 本研究方法 | IEEE 1609.2.1-2022 |
|---|---|---|
| 80 | 12.337 (46.223) | 18172.760 (5672.441) |
| 112 | 11.724 (19.598) | 32261.409 (14346.036) |
| 128 | 9.718 (85.111) | 21711.568 (2651.270) |
| 192 | 9.738 (57.772) | 19833.719 (2378.780) |
| 256 | 10.640 (26.082) | 24186.584 (5284.956) |

### 4.2.2 實驗 3 和實驗 4：擴展成 20 個繭公鑰、20 個蝴蝶公鑰

本節將整理實驗 3 和實驗 4 的實驗結果，觀察從毛蟲公鑰擴展成繭公鑰和從繭公鑰擴展成蝴蝶公鑰各擴展 20 個公鑰的結果，並且統計每個實驗中的 1,000 資料用“平均值(標準差)”的方式呈現每個方法的金鑰擴展時間，單位採用微秒。

實驗 3 的金鑰擴展時間整理如表 4 所示。其中，安全強度 80 時，本研究方法花費的金鑰擴展平均時間為 13.148 微秒，而 IEEE 1609.2.1-2022 方法花費的金鑰擴展平均時間為 37656.958 微秒。由於本研究方法在擴展成 1 個繭公鑰或擴展成 20 個繭公鑰時，主要只是做一般整數計算，所以實驗 1 和實驗 3 金鑰擴展時間沒有顯著差異。然而，在 IEEE 1609.2.1-2022 方法在擴展成 1 個繭公鑰或擴展成 20 個繭公鑰時，有部分參數是共用的或增量遞增的，所以如果使用的函式庫中有支援快速求冪演算法的話，將可以提升運算效率；因此，IEEE 1609.2.1-2022 方法在實驗 3 的每個繭公鑰金鑰擴展平均時間略有下降。

實驗 4 的金鑰擴展時間整理如表 5 所示。其中，安全強度 80 時，本研究方法花費的金鑰擴展平均時間為 13.202 微秒，而 IEEE 1609.2.1-2022 方法花費的金鑰擴展平均時間為 18684.688 微秒。由於本研究方法在擴展成 1 個蝴蝶公鑰或擴

展成 20 個蝴蝶公鑰時，主要只是做一般整數計算，所以實驗 2 和實驗 4 金鑰擴展時間沒有顯著差異。然而，在 IEEE 1609.2.1-2022 方法在擴展成 1 個蝴蝶公鑰或擴展成 20 個蝴蝶公鑰時，主要運算時間仍是花費在計算 $c$ 倍數的橢圓曲線基準點 $G$，所以實驗 2 和實驗 4 金鑰擴展時間沒有顯著差異。

**表 4　實驗 3 金鑰擴展時間(單位：微秒)**

| 安全強度 | 本研究方法 | **IEEE 1609.2.1-2022** |
|---|---|---|
| 80 | 13.148 (18.943) | 37656.958 (4003.968) |
| 112 | 12.393 (26.865) | 38377.835 (4676.397) |
| 128 | 12.845 (20.057) | 46556.214 (5849.735) |
| 192 | 10.050 (24.514) | 53909.124 (11794.927) |
| 256 | 10.335 (7.402) | 57894.552 (10633.871) |

**表 5　實驗 4 金鑰擴展時間(單位：微秒)**

| 安全強度 | 本研究方法 | **IEEE 1609.2.1-2022** |
|---|---|---|
| 80 | 13.202 (15.895) | 18684.688 (4945.681) |
| 112 | 11.198 (10.894) | 25290.976 (12080.073) |
| 128 | 10.345 (107.860) | 21381.615 (9460.058) |
| 192 | 9.057 (45.968) | 23523.532 (6527.510) |
| 256 | 9.733 (26.814) | 28129.762 (5538.829) |

### 4.2.3 小結與討論

本節假設本研究方法金鑰擴展平均時間為 $T_R$、IEEE 1609.2.1-2022 方法金鑰擴展平均時間為 $T_E$，通過 $T_E/T_R$ 可以計算出本研究方法與 IEEE 1609.2.1-2022 方法效率提升倍數。表 6 整理各個安全強度和各個實驗中，本研究方法與 IEEE 1609.2.1-2022 方法效率提升倍數。由實驗結果顯示，在各個不同的安全強度和實驗情境下，本研究方法相較於 IEEE 1609.2.1-2022 方法在金鑰擴展效率都有顯著提升，提升倍數從 1415 倍~5871 倍。

然而，因為論文篇幅限制，本研究僅比較和討論金鑰擴展效率，未來仍需對金鑰產製、加解密及簽驗章上的效率進行深入比較。其中，由於 RSA 演算法本身在金鑰產製時間將可能高於 ECC 演算法的金鑰產製時間。雖然本研究方法有

在金鑰產製時間的限制，但在金鑰產製主要在終端設備執行，而公鑰擴展則主要在註冊中心和授權憑證中心執行。以台灣車聯網的實際應用場域中，假設全台灣終端設備(包含車載設備(On-Board Unit, OBU)和路側設備(Road-Site Unit))有可能幾百萬台，而註冊中心和授權憑證中心僅有幾十台。因此，終端設備的數量級是註冊中心和授權憑證中心幾萬倍，並且金鑰產製是由各個終端設備各自執行，可以平行運算，不會造成整體車聯網安全憑證管理系統的系統瓶頸。在此情境下，註冊中心和授權憑證中心如果為每個終端設備各別產製數十個繭公鑰和蝴蝶公鑰的話，則有數千萬或億級的金鑰擴展需求，所以提升金鑰擴展效率仍是較重要的議題。

**表 6　本研究方法與 IEEE 1609.2.1-2022 方法效率提升倍數**

| 安全強度 | 實驗 1 | 實驗 2 | 實驗 3 | 實驗 4 |
|---|---|---|---|---|
| 80 | 4687 | 1473 | 2864 | 1415 |
| 112 | 4728 | 2752 | 3097 | 2258 |
| 128 | 4178 | 2234 | 3625 | 2067 |
| 192 | 5871 | 2037 | 5364 | 2597 |
| 256 | 5397 | 2273 | 5602 | 2890 |

# 5. 結論與未來研究

有鑑於車輛的隱私和產製假名憑證(即蝴蝶公鑰)效率的需求，本研究提出原創的高效率金鑰擴展方法，基於 RSA 密碼學建構另一套蝴蝶金鑰擴展方法，並且通過數學原理和實作證明本研究方法的安全性和效率。其中，在美國國家標準暨技術研究院所規範的安全強度標準(Barker, 2020)五種等級條件下，分別比較本研究方法和 IEEE 1609.2.1 在 2022 年版的方法，由實驗結果顯示本研究方法可以提升金鑰擴展效率 1415 倍~5871 倍。

本研究中提出的高效率金鑰擴展方法，雖然可以提供較高效率的金鑰擴展，但此方法在金鑰產製、加解密及簽驗章上的效率仍需深入探討和驗證。除此之外，在量子電腦問世後，RSA 基礎和 ECC 基礎的金鑰對都有可能被量子計算所破解，所以未來應朝向後量子密碼學(Post-Quantum Cryptography, PQC)的金鑰擴展方法來發展。

# 參考文獻

Ahmad, S.A., Hajisami, A., Krishnan, H., Ahmed-Zaid, F., Moradi-Pari, E. (2019). V2V system congestion control validation and performance. *IEEE Transactions on Vehicular Technology*, *68*(3), 2102-2110.

Barker, E. (2020). *Recommendation for key management: part 1 – general* (SP 800-57 Part 1 Rev. 5). National Institute of Standards and Technology. Access on 2022/12/16, https://csrc.nist.gov/publications/detail/sp/800-57-part-1/rev-5/final

Barreto, P.S.L.M., Simplicio, M.A., Ricardini, J.E., Patil, H.K. (2021). Schnorr-based implicit certification: improving the security and efficiency of vehicular communications. *IEEE Transactions on Computers*, *70*(3), 393-399.

Berlato, S., Centenaro, M., Ranise, S., (2022). Smart card-based identity management protocols for V2V and V2I communications in CCAM: a systematic literature review. *IEEE Transactions on Intelligent Transportation Systems*, *23*(8), 10086-10103.

Chen, L., Moody, D., Regenscheid, A., & Randall, K. (2019). *Recommendations for discrete logarithm-based cryptography: elliptic curve domain parameters* (SP 800-186 (Draft)). National Institute of Standards and Technology. Access on 2022/12/16, https://csrc.nist.gov/publications/detail/sp/800-186/draft

Directorate-General for Mobility and Transport (2018), *Certificate policy for deployment and operation of European cooperative intelligent transport systems (C-ITS)*(Release 1.1, June 2018). European Commission. Access on 2022/12/16, https://transport.ec.europa.eu/system/files/2018-05/c-its_certificate_policy-v1.1.pdf

Ghosal, A., & Conti, M. (2020). Security issues and challenges in V2X: a survey. Computer Networks. *169*, 107093.

Hammi, B., Monteuuis, J.-P., & Petit, J. (2022). PKIs in C-ITS: Security functions, architectures and projects: A survey. *Vehicular Communications*. *38*, 100531.

Intelligent Transportation Systems Committee. (2022). *IEEE standard for wireless access in vehicular environments (WAVE) - certificate management interfaces for end entities* (IEEE Std 1609.2.1™-2022). IEEE Vehicular Technology Society. Access on 2022/12/16, https://standards.ieee.org/ieee/1609.2.1/10728/

Molina-Masegosa, R., Sepulcre, M., Gozalvez, J., Berens, F., Martinez, V. (2020). Empirical models for the realistic generation of cooperative awareness messages in vehicular networks. *IEEE Transactions on Vehicular Technology*, *69*(5), 5713-5717.

Simplicio, M. A., Cominetti, E. L., Patil, H. K., Ricardini, J. E., & Silva, M. V. M. (2018). The unified butterfly effect: efficient security credential management system for vehicular communications. In the Proceedings of 2018 IEEE Vehicular Networking Conference (VNC), Taipei, Taiwan.

Walker, J. (2022). *Security credential management system (SCMS)*. U.S. Department of Transportation. Access on 2022/12/16, https://www.its.dot.gov/resources/scms.htm